\documentclass{aastex631}

\newcommand{\ztfrest}{ZTFReST}
\newcommand{\nmma}{\text{NMMA}}

\graphicspath{{./}{figures/}}

\usepackage{longtable}

\begin{document}

\title{IIb or not IIb: A Catalog of ZTF Kilonova Imposters}

\author[0000-0002-4843-345X]{Tyler Barna}
\affiliation{School of Physics and Astronomy, University of Minnesota, Minneapolis, Minnesota 55455, USA}

\author[0000-0002-4223-103X]{Christoffer Fremling}
\affiliation{Caltech Optical Observatories, California Institute of Technology, Pasadena, CA 91125, USA}
\affiliation{Division of Physics, Mathematics and Astronomy, California Institute of Technology, Pasadena, CA 91125, USA}

\author[0000-0002-2184-6430]{Tomas Ahumada}
\affiliation{Cahill Center for Astrophysics, California Institute of Technology, Pasadena, CA 91125, USA}

\author[0000-0002-8977-1498]{Igor Andreoni}
\affiliation{Department of Physics and Astronomy, University of North Carolina, Chapel Hill, NC 27599, USA}

\author[0000-0001-6595-2238]{Smaranika Banerjee}
\affiliation{The Oskar Klein Centre \& Department of Astronomy, Stockholm University, AlbaNova, SE-106 91 Stockholm, Sweden}

\author[0000-0002-7777-216X]{Joshua S. Bloom}
\affiliation{Department of Astronomy, University of California, Berkeley, CA 94720}
\affiliation{Physics Division, Lawrence Berkeley National Laboratory, 1 Cyclotron Road, MS 50B-4206, Berkeley, CA 94720, USA}

\author[0000-0002-8255-5127]{Mattia Bulla}
\affiliation{Department of Physics and Earth Science, University of Ferrara, via Saragat 1, I-44122 Ferrara, Italy}
\affiliation{INFN, Sezione di Ferrara, via Saragat 1, I-44122 Ferrara, Italy}
\affiliation{INAF, Osservatorio Astronomico d’Abruzzo, via Mentore Maggini snc, 64100 Teramo, Italy}

\author[0000-0001-9152-6224]{Tracy X. Chen} 
\affiliation{IPAC, California Institute of Technology, 1200 E. California Blvd, Pasadena, CA 91125, USA}

\author[0000-0002-8262-2924]{Michael W. Coughlin}
\affiliation{School of Physics and Astronomy, University of Minnesota, Minneapolis, Minnesota 55455, USA}

\author[0000-0003-2374-307X]{Tim Dietrich}
\affiliation{Institute of Physics and Astronomy, Theoretical Astrophysics, University Potsdam, Haus 28, Karl-Liebknecht-Str. 24/25, 14476, Potsdam, Germany}
\affiliation{Max Planck Institute for Gravitational Physics (Albert Einstein Institute), Am Mühlenberg 1, Potsdam 14476, Germany}

\author[0000-0002-9364-5419]{Xander J. Hall}
\affiliation{McWilliams Center for Cosmology and Astrophysics, Department of Physics, Carnegie Mellon University, 5000 Forbes Avenue, Pittsburgh, PA 15213}

\author[0000-0002-9380-7983]{Alexandra Junell}
\affiliation{School of Physics and Astronomy, University of Minnesota, Minneapolis, Minnesota 55455, USA}

\author[0000-0001-7648-4142]{Ben Rusholme} 
\affiliation{IPAC, California Institute of Technology, 1200 E. California Blvd, Pasadena, CA 91125, USA}

\author[0000-0003-1546-6615]{Jesper Sollerman}
\affiliation{The Oskar Klein Centre \& Department of Astronomy, Stockholm University, AlbaNova, SE-106 91 Stockholm, Sweden}

\author{Niharika Sravan}
\affiliation{Cahill Center for Astrophysics, California Institute of Technology, Pasadena, CA 91125, USA}
\affiliation{Department of Physics, Drexel University, Philadelphia, PA 19104, USA}


\begin{abstract}

Among the various classes of fast optical transients (FOTs), kilonovae (KNe), which can emerge as a result of neutron star mergers, are extremely challenging to observe because of not only the rapid timescale on which they fade (on the order of days), but also due to the relative scarcity of their occurrence. This scarcity is compounded by the large number of other FOTs that may initially resemble the characteristic rise of a KNe. While these objects can be ruled out as candidate KNe by taking spectroscopy, a method of confidently ruling out transients based on photometric analysis alone would be incredibly valuable. We describe the compilation of various ``imposter" transients, including a plurality of IIb SNe, and investigate a number of comparative metrics by which one might be able to remove transients from consideration without the use of spectroscopy. We provide a list of these objects and their classifications as well as a glossary of the transient types included in the sample.

\end{abstract}

\section{Introduction} \label{sec:intro}

Kilonovae, which are powered by the thermal heating caused by the radioactive decay of newly synthesized and unstable \textit{r}-elements \citep{1998ApJ...507L..59L, 2019LRR....23....1M}, can provide smoking gun evidence of neutron star mergers. This thermal emission peaks around a day after the initial merger in the optical regime, followed by a rapid fade on the order of less than a week \citep{2010MNRAS.406.2650M}, with longer wavelength emissions demonstrating a later peak and more gradual fade \citep{2017ApJ...851L..21V, 2013ApJ...775..113T}. In addition to the aforementioned electromagnetic (EM) signal, these mergers can also produce a $\gamma$-ray burst (GRB) \citep{2013Natur.500..547T, 2013ApJ...774L..23B, 2013ApJ...776...18F, 2015NatCo...6.7323Y} and are detectable via gravitational waves (GW) generated from the compact binary coalescence (CBC) \citep{2015IJMPD..2430012R, 2020PhR...886....1N, 2021JPlPh..87a8402A}. It was through detections by the International Gravitational Wave Network (IGWN) of the binary neutron star merger GW170817  \citep{2017PhRvL.119p1101A}
that the first (and only) KN associated with a GW signal, AT2017gfo, was localized and subsequently observed in a spate of multi-wavelength observations \citep{CoFo2017,SmCh2017,AbEA2017f} in addition to the associated GRB170817A \citep{GoVe2017,SaFe2017,AbEA2017e}. 

Given the significant scientific value of KNe observations, a great deal of work has been dedicated to optimizing the search for these FOTs. In particular, work is being done to more effectively search for KNe in untargeted surveys, such as the Zwicky Transient Facility (ZTF) \citep{2019PASP..131a8002B, 2019PASP..131a8003M, 2019PASP..131g8001G, 2019PASP..131f8003B, 2020PASP..132c8001D}. These ``serendipitous" observations are an attractive prospect as they are not reliant on an associated GW signal, but this approach presents its own challenges. 
Some of these efforts are discussed in more detail in Section \ref{sec:methods}. Due to the characteristic rapid fade of KNe, observation has proven to be challenging \citep{2019MNRAS.486..672A, 2020ApJ...905..145K, 2024PASP..136k4201A}. Another difficulty faced by observers is the sheer number of FOTs that can initially display characteristics in line with what would be expected from a serendipitous KN observation. 

In order to determine whether or not these candidates are ``imposters," spectroscopy has to be taken on a rapid timescale. This is problematic due to the relatively limited rapid spectroscopic resources available to astronomers in comparison to photometric target of opportunity (ToO) resources. In the future, this will become an even greater challenge; the Vera C. Rubin Observatory's Legacy Survey of Space and Time (LSST) will produce a volume of data orders of magnitude greater than prior surveys, observing far more transients than can be reasonably followed up and classified using spectroscopy \citep{2019ApJ...873..111I}.


In this paper, we compile a sample of transients observed by ZTF over the course of 6 years that were at one point suspected to be KNe but were later determined to be imposters. Section~\ref{sec:methods} describes some of the tools and pipelines developed by astronomers to consolidate the number of transients that need to be investigated as candidate KNe. Section~\ref{sec:candidates} describes the sample of transients in greater detail, including the distribution of transient types; this is followed by a description of how the lightcurves of these objects are interpolated for more consistent analysis in Section~\ref{sec:DP}. We evaluate a number of metrics derived from these lightcurves and compare them to a KN model grid in Section~\ref{sec:eval}. The results of these evaluations and commentary on the potential for additional metrics are discussed in Section~\ref{sec:conclusions}.

\section{Methods}\label{sec:methods}
The detection of GW170817 and AT2017gfo promptly motivated developments in automating searches for KNe \citep{2017PASA...34...37A}, which led to the discovery of a number of FOTs. A large number of the transients in our sample were initially identified as potential KNe by the framework described below in Section~\ref{sec:ztfrest}.

\subsection{ZTFReST}\label{sec:ztfrest}
The ZTF Realtime Search and Triggering (\ztfrest) framework was developed following the conclusion of the O3 IGWN observing run \citep{2021ApJ...918...63A}. Rather than searching for KNe by waiting for GRB or GW localizations, \ztfrest \ uses ZTF to search for KNe and FOTs that might be serendipitously observed in its wide-field survey. To accomplish this, it must employ a number of strategies to reduce the dataset, which consists of upwards of a million alerts per night.
By applying a requirement that objects fade at a rate of at least 0.3 magnitudes per day, \ztfrest \ identifies rapidly fading objects\footnote{Readers may notice that some of the objects in our sample do not exceed the fade rate imposed by \ztfrest. This is due to multiple factors; the objects in our sample are not exclusively sourced from the \ztfrest \ pipeline. Additionally, the rate calculated by \ztfrest \ is based on the difference between individual points of observation at the time they occur - while some of the objects may have exceeded the \ztfrest \ rate based on two data points, their average fade rate over the course of two days following the peak may be lower.}, which reduces the number of nightly objects considerably. There are several additional processing steps made as part of the \ztfrest \ pipeline, which are summarized in Figure 1 of \cite{2021ApJ...918...63A}.
 
While \ztfrest \ has not yet detected a KN, it has been successful in identifying a number of interesting FOTs, including several GRB afterglows \citep{2021ApJ...918...63A} and a jetted tidal disruption event (TDE) \citep{2022Natur.612..430A}. Additionally, \cite{2021ApJ...918...63A} constrained the rate of GW170817-like KNe to $R<900 \text{ Gpc}^{-3}\text{yr}^{-1}$.

\subsection{NMMA}\label{sec:nmma}

The Nuclear Multi Messenger Astrophysics (NMMA) Framework (\nmma)~\citep{2020Sci...370.1450D, 2023NatCo..14.8352P} is a Bayesian inference tool for analyzing astrophysical transients using EM and GW signals together with information obtained from other EM observations or from nuclear-physics theory and experiments~\citep{Koehn:2024set}. While the correct inference of the source properties, such as the component masses of neutron star mergers or the extraction of information to further sharpen our knowledge about the equation of state of supranuclear-dense matter, e.g.,~\citep{2020Sci...370.1450D,Pang:2021jta,Pang:2023dqj,Koehn:2024ape}, is among the main science goals of NMMA, its usage to perform model selection and to classify EM transients, e.g., \citep{Kunert:2023vqd,Kann:2023ulv} is an ability that is of greatest interest for this paper. 

For this purpose, NMMA allows the comparison of numerous EM transient models, e.g., for the description of different kinds of KNe caused by binary neutron star or black hole - neutron star mergers~\citep{Kasen:2017sxr,Bulla:2019muo,Bulla:2022mwo}, for the description of collapsars~\citep{Barnes:2022dxv,Barnes:2023ixp} and supernovae~\citep{Levan:2004sn}, or for the modeling of GRB afterglows~\citep{Ryan:2019fhz}. These different models can be compared against the observational data through a Bayesian Inference Analysis, and by comparing the obtained Bayes factor, one can quantitatively predict the relative changes that the observed FOT was better described by the employed KN models compared to other scenarios. 
Along these lines, \cite{10.1093/mnras/stae1164} discussed an online framework that uses \nmma \ to analyze objects that meet the criteria of \ztfrest , fitting objects to several different transient models, quantifying the chances of a KN detection. 

\section{Candidates}\label{sec:candidates}

\begin{figure}
    \centering
    \includegraphics[width=0.5\columnwidth]{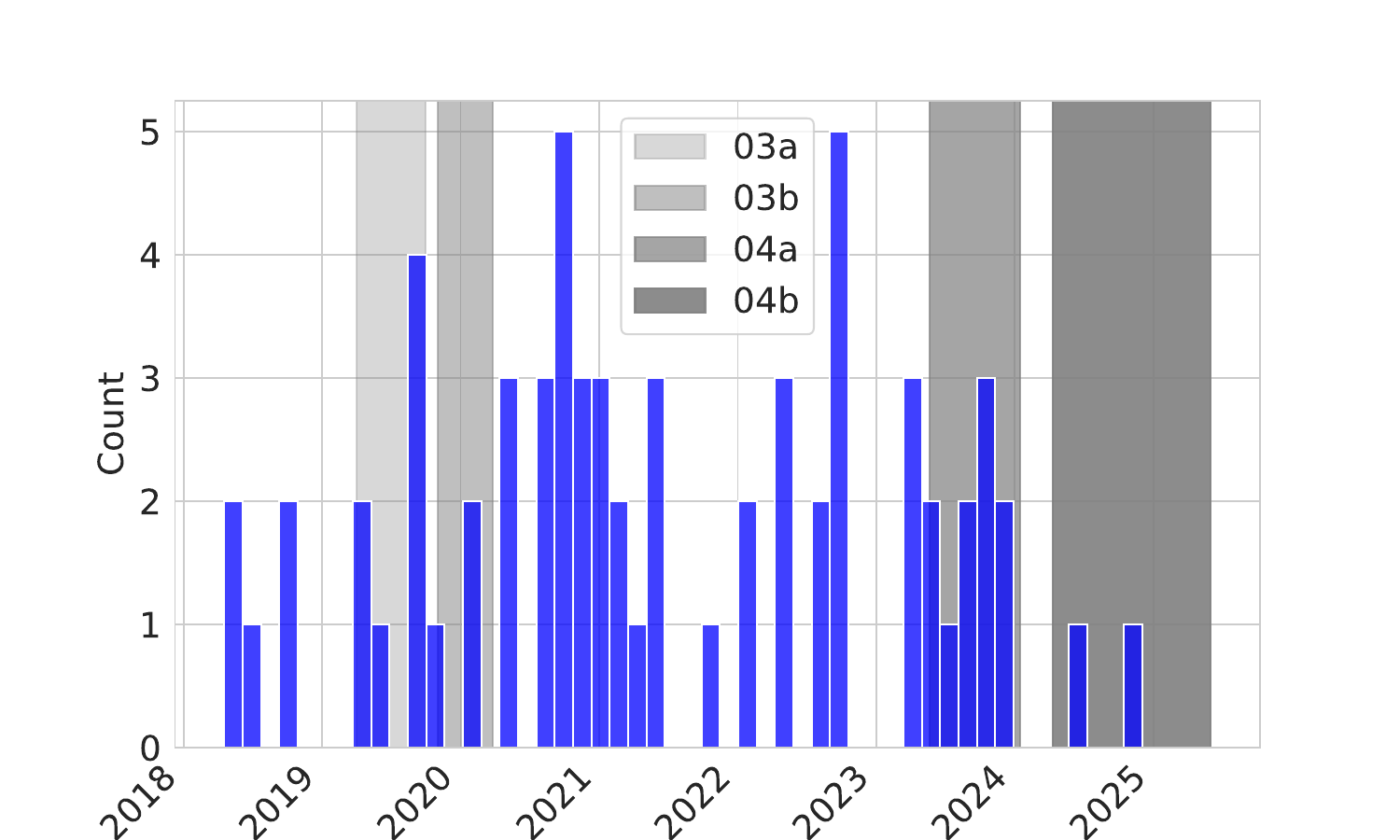}
    \caption{Timeline of observations in the transient sample as compared to the most recent LIGO observing runs. The 04 observing run is set to conclude at the end of 2025, with 05 beginning sometime in mid-2027.}
    \label{fig:kncandidates}
\end{figure}

The initial list of compiled objects was comprised of 192 transients observed between 2018 and 2024 by ZTF. Each of these objects was initially flagged as being a fast transient by ZTFReST and associated fast transient monitoring projects. 
From these 192 transients, only 86 were investigated, formally classified, and reported to the Transient Name Server, generally based on spectroscopy. Since our study primarily concentrates on identifying any trends in specific transient classes with respect to the classification of KNe, we restrict ourselves to the 86 as our final sample set. A subset of these objects was observed during IGWN observing runs, which we highlight in Figure \ref{fig:kncandidates}.

An additional requirement of objects with known redshifts lowers the number of transients to 70. This includes 12 different classes of transient, with the majority of the sample being SNe IIb at 20 unique objects. There is an exception to the redshift requirement for the galactic sources (cataclysmic variables and novae). This is because many of these objects have a reported redshift of 0 due to the difficulty of deriving the redshift for galactic sources from spectroscopy. For these objects, we assume a distance equal to the center of the galaxy, $8178 \ \text{pc}$ \citep{2019A&A...625L..10G}. This results in absolute magnitudes for "zero redshift" sources roughly in line with the handful of galactic sources with measured redshifts as well as analyses in the literature \citep{2010JAVSO..38..193K, 2019A&A...622A.186S}. As detailed in Section \ref{sec:GP}, an additional 4 objects are not suited to the Gaussian Process (GP) method, so they are also excluded. This brings the total number of objects in the data set to 66.

\begin{figure}
    \centering
    \includegraphics[width=0.44\columnwidth]{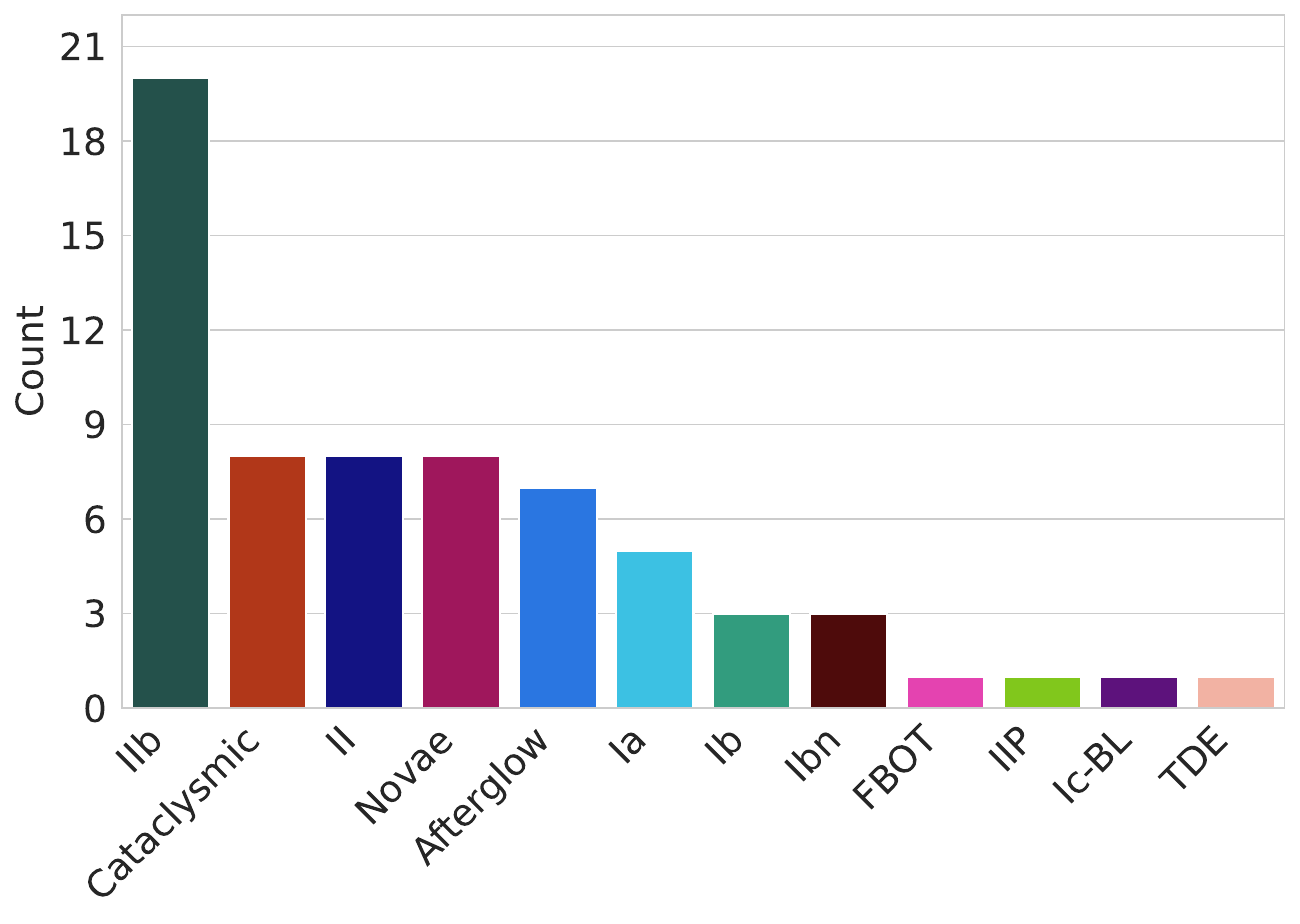}
    \caption{The distribution of object types included within the sample. The largest transient class in the sample is SNe IIb.}
    \label{fig:type-histogram}
\end{figure}

\section{Data Processing}\label{sec:DP}

In the aggregate, all of the transient types in this study are characterized by a rise followed by a fall. Due to the nature of transient observation cadences, a smaller number of objects were observed during their rise compared to the number observed during their fade. Of the objects that passed the GP quality requirement, there are 3 objects observed only during their rise, 33 objects observed only during their fade, and 30 objects observed for both their rise and fade. The distribution of object types in the sample is shown in Figure \ref{fig:type-histogram}. A brief description of each of these classes can be found in Appendix \ref{sec:classes}.

IIb SNe, which make up the plurality of the sample, are core-collapse SNe characterized by a double peak separated by several days \citep{2011A&A...528A.131C}. The first peak occurs as a result of shock breakout and demonstrates a rapid rise and fade on the order of a few days. As this shock expands and becomes optically thin, $^{56}$Ni produced during the collapse begins to decay, heating the ejecta and giving rise to a second peak. The second peak is followed by a much more gradual fall than the first \citep{1994ApJ...429..300W}.

This double peak is not observed in all of the IIb SNe in our sample; rather, some IIb SNe in the sample were only observed for the former or the latter. In cases where there are two peaks present (or a clear fade prior to a second rise), the first peak is used for the calculation of the rise and fade rates. This is relevant to consider when analyzing some of the metrics discussed in Section \ref{sec:eval}.

Observational data is cut to center around the rise, peak, and fade. Of the 33 objects with an observed rise, 8 have a rise duration that exceeds 10 days in any filter; of these, 5 are SNe IIb and 3 are SNe Ia. For the purposes of this paper, we measure the fade as the difference in magnitude between the peak luminosity and observations over the next 2 days. KNe, as fast fading objects, would generally only be visible for a few days following their peak. An object that is still observable by ZTF for weeks following its peak is unlikely to be a KN and can be excluded from consideration, but the longevity of a transient cannot always be ascertained from a handful of early-time observations. 

\subsection{Gaussian Process}\label{sec:GP}
Because this sample is made up of a mix of both untargeted and ToO observations, the observation cadence is not consistent. Limiting all objects to data collected within the first 25 days of observation, objects were observed a median of 22 times at an average cadence of 1.38, 1.71, and 3.68 days in the \textit{g}-band, \textit{r}-band, and \textit{i}-band, respectively. As a result, observations are not evenly distributed across the three filters. \textit{g}-band and \textit{r}-band observations make up 47\% and 46\% of the observed data points, respectively, whereas the \textit{i}-band accounts for only 7\% of observations. Additionally, the rapid nature of these transients in both their rise and fade means that calculating rates purely based on the observed data points is challenging. 

In order to analyze the rise and fade of these transients, it is necessary to interpolate them on a more uniform observing cadence. Due to the irregular nature of observing cadences and the limited time range in consideration for most of these objects, linear interpolation proved insufficient for the purposes of this study. Instead, we applied a GP method to the objects. 

In this paper, we use a pre-trained GP kernel, which was trained on approximately 14,000 objects observed by ZTF. This training set contains a number of objects, with 60.6\% of the sample being made up of Type I and Type II SNe. All observations and their associated errors in the dataset are converted to flux

\begin{equation}\label{eqn:flux}
    F = 10^{-0.4 (m_{AB} - 23.9)} \mu \text{Jy},
\end{equation}

\begin{equation}\label{eqn:flux-error}
    \sigma_F = \left( \frac{\sigma_m}{2.5 / \ln(10)} \right) \cdot F.
\end{equation}
where $\sigma_m$ and $\sigma_F$ are the magnitude and flux errors, respectively.

Using these flux values, a GP kernel is trained using \texttt{george} \citep{2015ITPAM..38..252A}. From this kernel, it is possible to interpolate ZTF lightcurves to estimate high cadence observations.

\begin{figure}[h]
    \centering
    \includegraphics[width=0.6\linewidth]{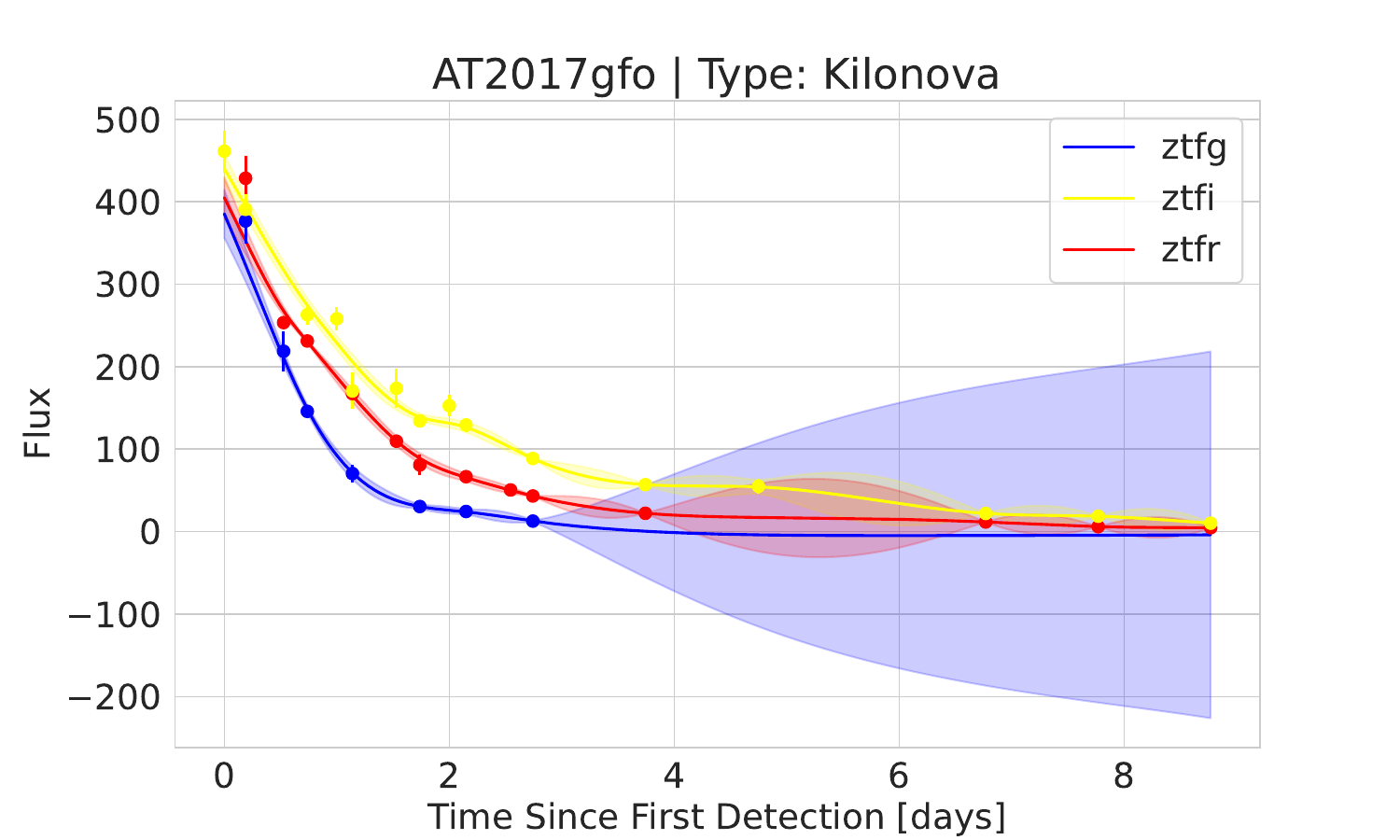}
    \caption{The GP as applied to AT2017gfo observations from Pan-STARRS. While the rise of the KN is not captured, the fade is well sampled and is modeled well by the GP.}
    \label{fig:AT2017gfo}
\end{figure}

The GP is applied to the first 50 days of data for all objects in our sample\footnote{We restrict our analysis to the first 50 days because this study largely focuses on the rapid rise and fade of transients, making longer term data unnecessary.}. This results in smooth light curves with high observing cadences, allowing for more precise analysis of rapid changes in luminosity. From there, each of the objects was reduced to only include observations at and around the peak luminosity, and the flux values from the GP are converted back to magnitude space by inverting Equations \ref{eqn:flux} and \ref{eqn:flux-error}. An example of the GP applied to AT 2017gfo is shown in Figure \ref{fig:AT2017gfo}. 

\begin{figure}[h]
    \centering
    \includegraphics[width=0.6\linewidth]{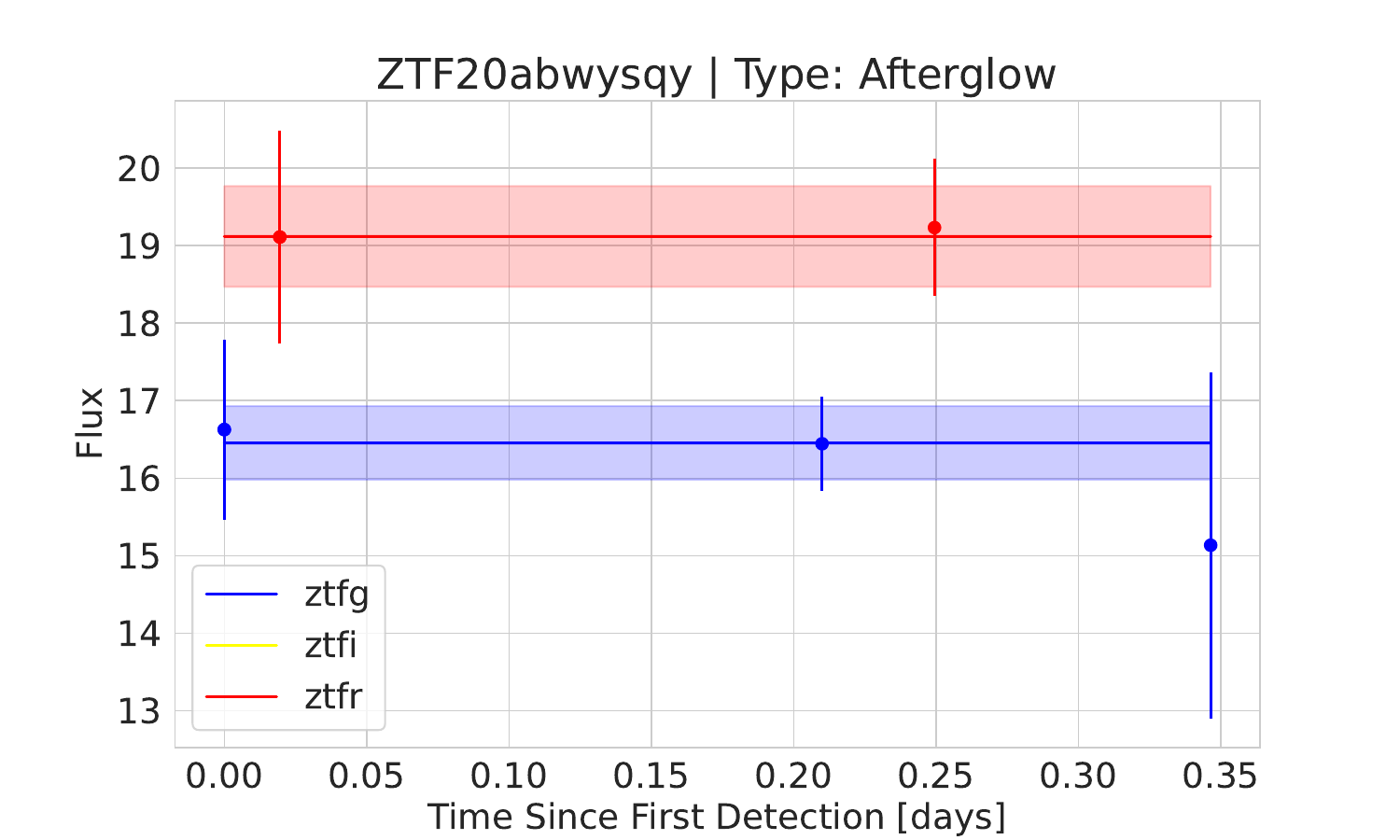}
    \caption{An example of a long GRB that the GP was unable to meaningfully fit. Observations are plotted as points with errors, with the line plot representing the GP. ZTF was only able to observe this object for a single night \citep{ZTF20abwysqy}.}
    \label{fig:ZTF20abwysqy}
\end{figure}

After applying the GP to the 70 objects with classifications and redshifts, 4 of the transients did not have sufficient data to be modeled by the GP; an example of one such object can be seen in Figure \ref{fig:ZTF20abwysqy}, which the GP models as essentially flat. This is due to the relatively low number of observations, which take place over a single night. The third data point in the \textit{g}-band suggests there may be an observed fade, but the error is too large to say this definitively. Regardless, this shows that there is a lower limit in terms of observations below which the GP cannot effectively interpolate the data.

\begin{figure}[h]
    \centering
    \includegraphics[width=0.6\linewidth]{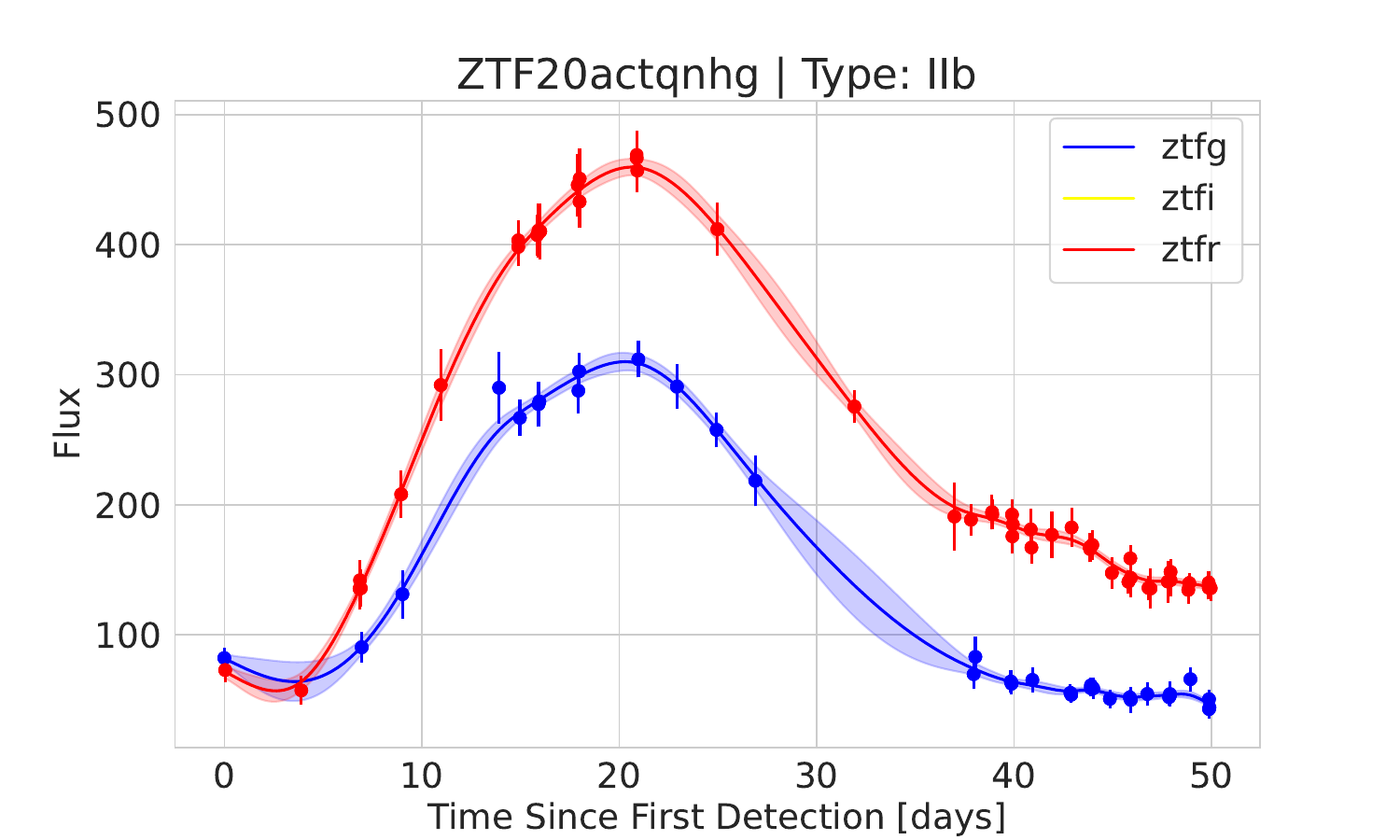}
    \caption{The GP as applied to a IIb SN; again, observations are plotted as points, with the lines showing the GP interpolation. This is a more plausible interpolation compared to ZTF20abwysqy}
    \label{fig:ZTF20actqnhg}
\end{figure}

Conversely, Figure \ref{fig:ZTF20actqnhg} shows a IIb SN, ZTF20actqnhg, that was well fit by the GP. While this is more observed than ZTF20abwysqy, this is an unevenly sampled lightcurve with a somewhat ambiguous peak in the \textit{g}-band and a paucity of observations immediately following the peak in the \textit{r}-band. These properties would make calculating several metrics, including the color at peak and FWHM, less consistent based solely on the observed data points. The GP creates a plausible interpolation of the observed data, which enables more consistent calculations of metrics like the fade rate and FWHM.

\begin{figure}[h]
    \centering
    \includegraphics[width=0.6\linewidth]{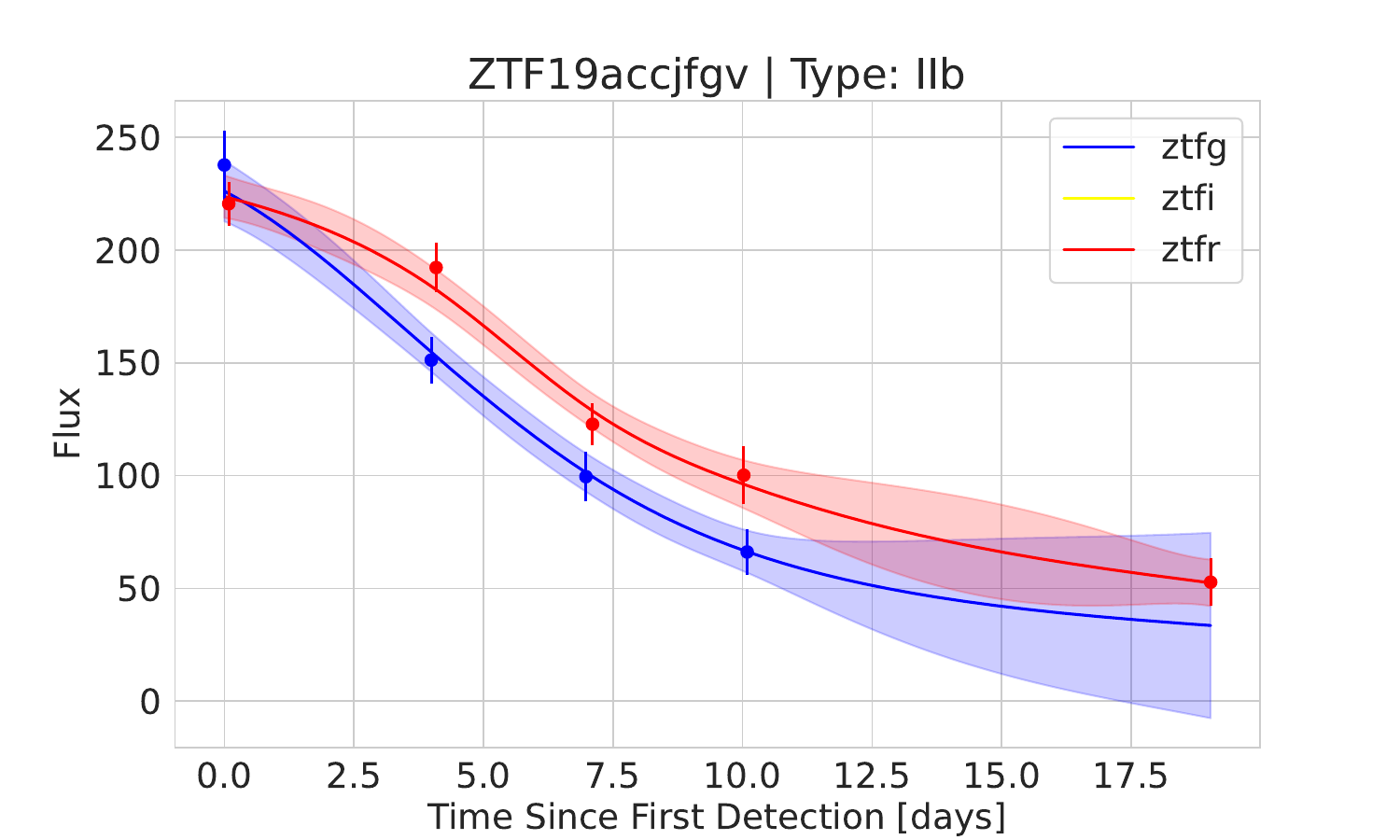}
    \caption{The GP as applied to an object without an observed rise. The cadence of observation would preclude the calculation of the fade rate of the two days following the peak observation, but it is possible to calculate with the GP interpolation.}
    \label{fig:ZTF19accjfgv}
\end{figure}

Figure \ref{fig:ZTF19accjfgv} shows ZTF19accjfgv, a SN IIb that was only observed while it was fading. There is a gap of around 3 days between the first and second observations in each filter. Recall from Section \ref{sec:DP} that we calculate the fade rate as the change in luminosity over the course of 2 days following the peak observation. Calculating the fade rate using only observed data points would therefore result in this object not being usable for this study. However, applying the GP results in a plausible interpolation that fits all observed points and allows us to evaluate the fade rate on a consistent time interval for all objects.

As a result, the total number of objects included in the study is 66, with 11 different transient types represented in the sample. The distribution of types in the sample is shown in Figure \ref{fig:type-histogram}. A full list of objects can be found in Table \ref{tab:objects}.

The benefit of using the GP to interpolate the lightcurves is most apparent in the \textit{i}-band, which is generally less available than \textit{g}-band and \textit{r}-band observations. The use of the GP results in 30 objects having smooth rise, fade, or both values in the \textit{i}-band.

\section{Evaluation Metrics}\label{sec:eval}

After applying the GP, we calculate a number of 
derived observational metrics for the transients. These include the rise, the fade, the color at each filter's peak, and the full width at half maximum (FWHM). The specifics of these calculations are described in further detail in their respective subsections below.

We compare the sample of transients to a grid of kilonova models computed with the 3D Monte Carlo radiative transfer code \texttt{POSSIS} \citep{Bul2019,Bul2023}. \texttt{POSSIS} can handle arbitrary 3D geometry for the ejecta and simulate KN flux and polarization spectra for different observer viewing angles. Thermal radiation is represented by Monte Carlo packets that are assigned initial energies and frequencies and interact with matter while traveling through the expanding and opaque ejecta. In particular, we use the grid presented in \citep{Dietrich2020}\footnote{The grid is publicly available at \url{https://bit.ly/possis_models}}. The ejecta are assumed to be axially symmetric and are divided into two main components: a disk-wind ejecta component in the inner region (i.e., low velocities) and a dynamical ejecta component in the outer regions (i.e., high velocities) characterized by lanthanide-rich compositions close to the merger plane and lanthanide-poor compositions at polar angles. The masses of the two components, $m_{\rm wind}\in[0.01,0.03,0.05,0.07,0.09,0.11,0.13]\,M_\odot$ and $m_{\rm dyn}\in[0.001,0.005,0.01,0.02]\,M_\odot$, and the half-opening angle of the lanthanide-rich component, $\phi\in[0,15,30,45,60,75,90]^\circ$, are varied to create a grid of 196 models, with observables extracted for $11$ different viewing angles. All models share the same velocity structure, with the wind ejecta extending from 0.025 to 0.08c and the dynamical ejecta from 0.08 to 0.3c. The values for the model grid are generally plotted as black points in all figures.

We also calculate these metrics for AT2017gfo and include it under the ``AT2017gfo" label. We note that, since ZTF was not operating at the time of its detection, there are no ZTF observations of AT2017gfo. Instead, we use observations from Pan-STARRS \citep{2016arXiv161205560C}, which captured the fade in 3 bands similar to those of ZTF\footnote{We chose to use Pan-STARRS data for AT2017gfo due to the number of observations, from which the GP benefits. We elected to not include two other prominent KNe asssociated with GRBs 211211A \citep{2022Natur.612..223R} and 230307A \citep{2023GCN.33569....1L, 2024Natur.626..742Y} in our analysis. In the former case, there are not enough Pan-STARRS observations to meaningfully fit with the GP, and in the latter case, it was not observed by Pan-STARRS.}. The same GP is applied to these observations, which can be seen in Figure \ref{fig:AT2017gfo}. Across all the metrics evaluated, AT2017gfo generally falls within the same range as the KN model grid.

In addition to the following sections, which consider a combination of metrics, we evaluate the numerical separation between transient classes and the model grid with respect to each metric using a "distance" method that is described in Section \ref{sec:metric-distance}.

\subsection{Rise and Fade Rates}
\begin{figure}
    \centering
    \includegraphics[width=1\linewidth]{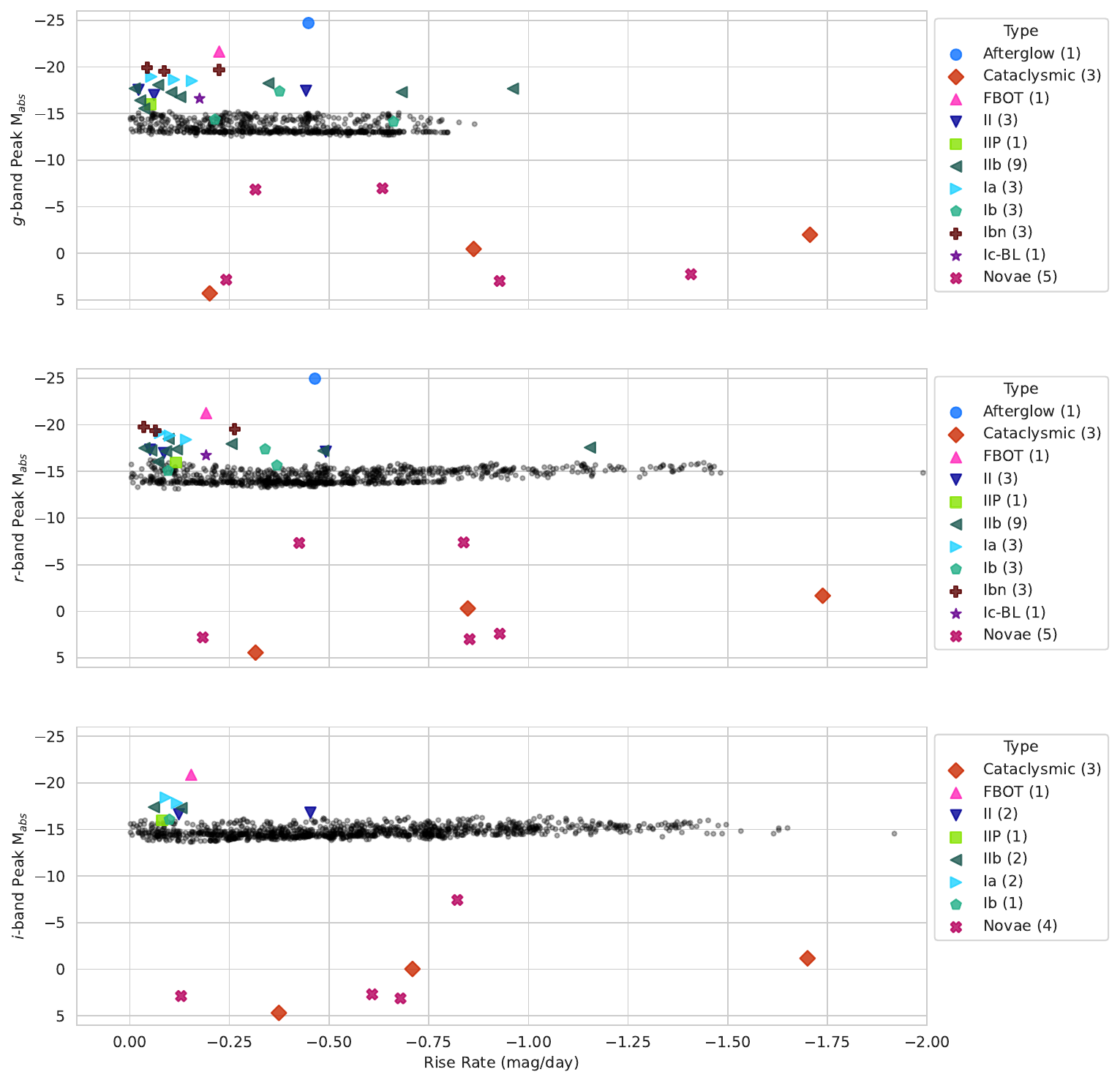}
    \caption{Rise rate vs.\ peak luminosity in each filter. Each subplot corresponds to a different filter, and there is a legend for each subplot with a count of each transient type present. Black points show the distribution of the KNe model grid.}
    \label{fig:rise-vs-peak}
\end{figure}

The rise rate is calculated by comparing the change in magnitude between an object's initial observation and the peak measured luminosity. While a handful of objects demonstrate longer rise times, the median (average) rise time is 2.96 (6) days.

Figure \ref{fig:rise-vs-peak} shows the peak magnitude in each filter versus the rise rate. The black points represent the KN model grid. While the transients in the cluster are somewhat clustered towards a lower rise rate than most of the lightcurves in the model grid in all filters, there are still a number of transients that have higher rise rates. There also exists a population of ``local" (z$<$0.001) objects with absolute magnitudes on the order of 5-15 magnitudes fainter than the model grid, with most of the remaining transient classes having absolute magnitudes similar to the model grid.

\begin{figure} 
    \centering
    \includegraphics[width=1\linewidth]{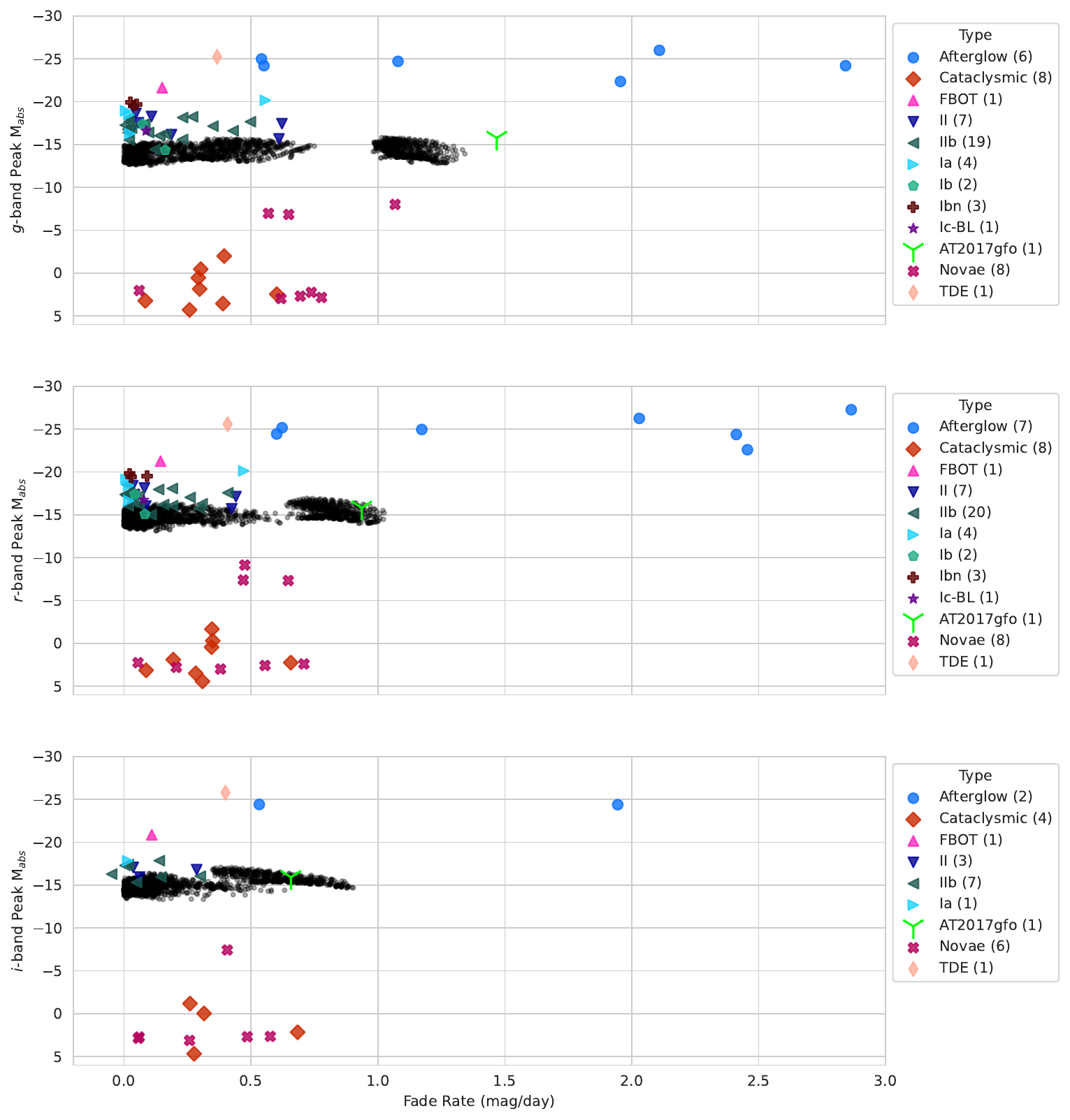}
    \caption{Fade rate vs/\ peak luminosity in each filter for the transient sample compared to the KNe model grid. Generally more objects were observed during their fade as compared to their rise (See Figure \ref{fig:rise-vs-peak}). Most of the transients are clustered just above the KNe model grid with comparable fade rates. The model grid demonstrates two populations distinguished by fade rate, with a gap between the faster and slower fading populations that is most noticeable in the \textit{g}-band. This bimodality is driven by a difference in the temperature parameter; the faster fading population is made up of simulated lightcurves with a lower temperature one day after the initial coalescence.}
    \label{fig:fade-vs-peak}
\end{figure}

Figure \ref{fig:fade-vs-peak} plots the fade rate versus the peak luminosity in each filter; the fade rate takes into account the change in magnitude between the time of peak luminosity and the following 2 days of data. This figure also shows a feature of the KN model grid. There are seemingly two populations of KNe, with a smaller subpopulation exhibiting a more rapid fade in the \textit{g}-band and, to a lesser extent, \textit{r}-band. There are several GRB afterglows that exceed the fade rate of the fast-fading KN population, which is to be expected \citep{andreoni2020constraining}. Most of the other transients are clustered towards the slower-fading population in all filters

\subsection{Color}
\begin{figure}
    \centering
    \includegraphics[width=1\linewidth]{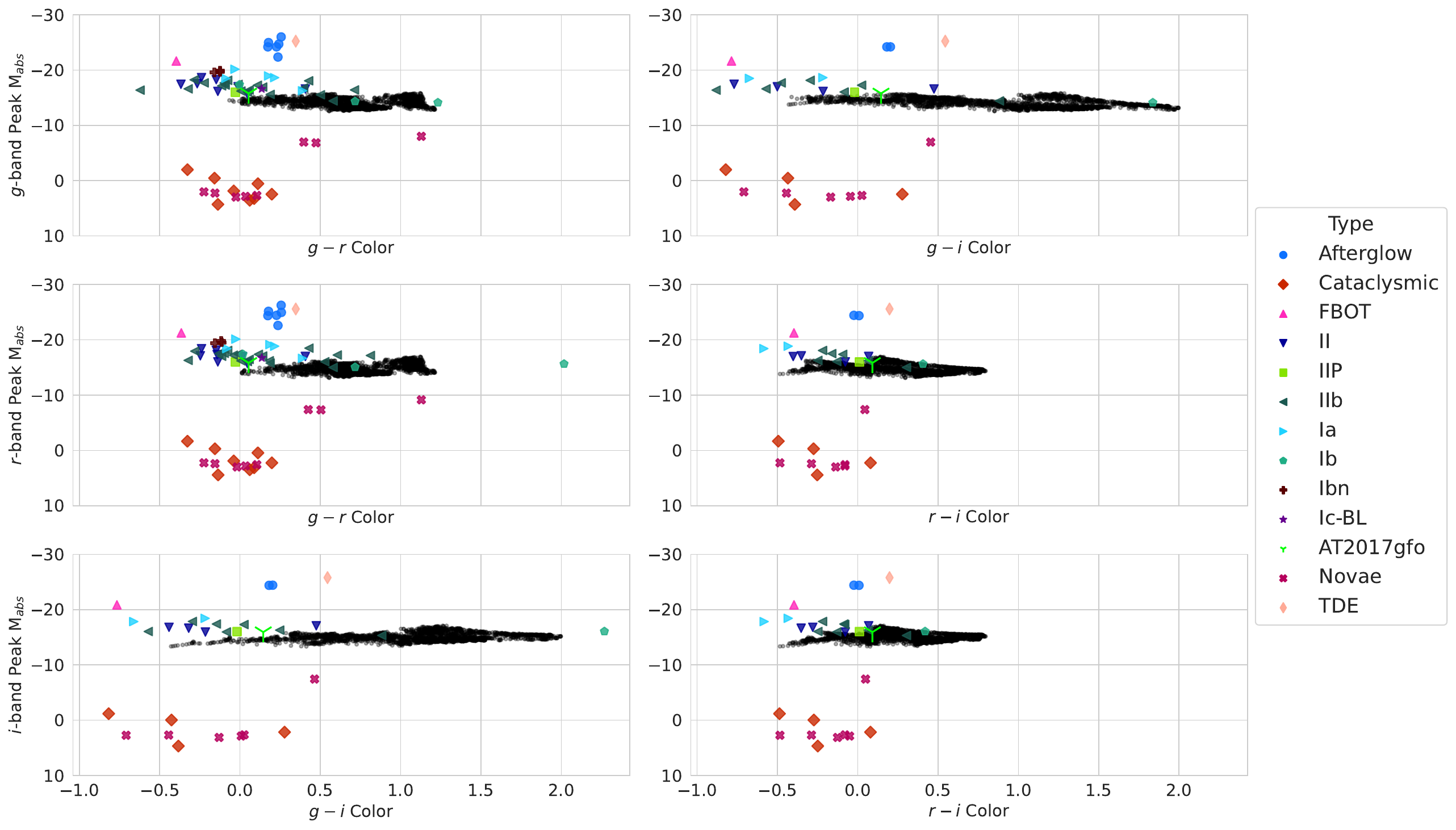}
    \caption{Color vs/\ Peak Luminosity in each filter. Each row plots the two color comparisons at the peak in the filter specified by the y-axis. Each subplot has the color being plotted labeled by their x-axis.}
    \label{fig:color-vs-peak}
\end{figure}

Figure \ref{fig:color-vs-peak} shows the color at the peak as compared to the peak luminosity in each filter. This metric seems to have a higher discriminating power than the rise or fade based on visual inspection, most notably with the KNe model grid being more red in the \textit{g-r} color at the \textit{g}-band peak than our sample of transients. The \textit{g-i} colors also show promise for both the \textit{g}-band and \textit{i}-band peaks.



\subsection{Full Width at Half Maximum}\label{sec:fwhm}
\begin{figure}
    \centering
    \includegraphics[width=0.9\linewidth]{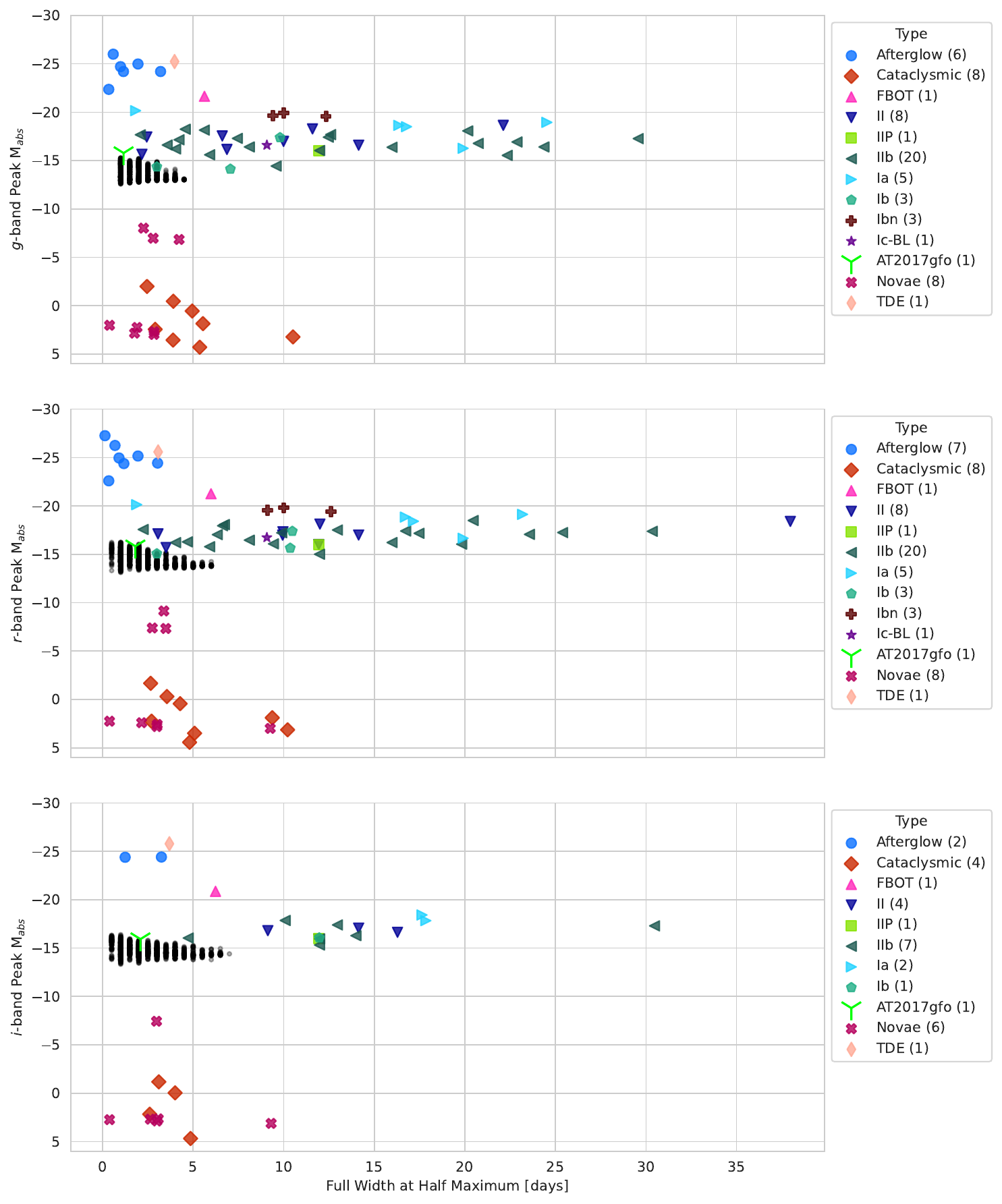}
    \caption{Time spent above half the peak luminosity vs the peak luminosity in each filter. There are a handful of transient classes that demonstrate FWHM values similar to the KN model grid, but many of the SNe demonstrate a much longer FWHM than the fast transients, which is generally due to the more rapid fading of KNe.}
    \label{fig:fwhm-vs-peak}
\end{figure}


The FWHM represents the duration an object remains above half of its maximum brightness while rising and fading.
This corresponds to a difference in luminosity of $-2.5 \ \text{log}\left( \frac{1}{2}\right) \simeq 0.75$ magnitudes.
Given that not all objects in the sample have both a rise and fade, objects with only an observed fade have their FWHM calculated as two times the duration they spend above half the maximum brightness while fading. This can be validated by comparing this same calculation for objects observed during their rise and fade against their true FWHM. Across all filters, this approximation overestimated the FWHM by between 10\%-20\%. However, a qualitative comparison of the objects using this estimation shows that they are within a similar range to the FWHM values that are not calculated with this estimation. This calculation is made for each filter a given object was observed in.

The FWHM is compared to the peak luminosity in Figure \ref{fig:fwhm-vs-peak}. This is perhaps the most promising of the metrics evaluated in this study. While there is still a population of transients that have similar FWHM values to the KN model grid, there are far more objects that lie outside the 0.5-7 day maximum FWHM shown by the KN grid. 
For example, in the \textit{g}-band, 14 of the 21 IIb SNe, 5 of the 8 II SNe, 4 of the 5 Ia SNe, and 3 of the 3 Ibn SNe had FWHM values that exceeded the KN model grid. The prevalence of IIb SNe with large FWHM values may be due to the observed peak being the second peak in their lightcurves, which tends to fade on a longer timescale in comparison to their first peak. We choose to include these IIb SNe to represent cases of imposters where the second peak is the only observed peak, which is not necessarily discernible with only a handful of data points following the peak.

That being said, this trend is not universal. In the same filter, most of the cataclysmic variables, novae, and classical novae have FWHM values in the same range as the KN model grid; this is also true of several other transient classes, but this could also be due to them only having 1-2 data points within this sample.

\begin{figure}
    \centering
    \includegraphics[width=1\linewidth]{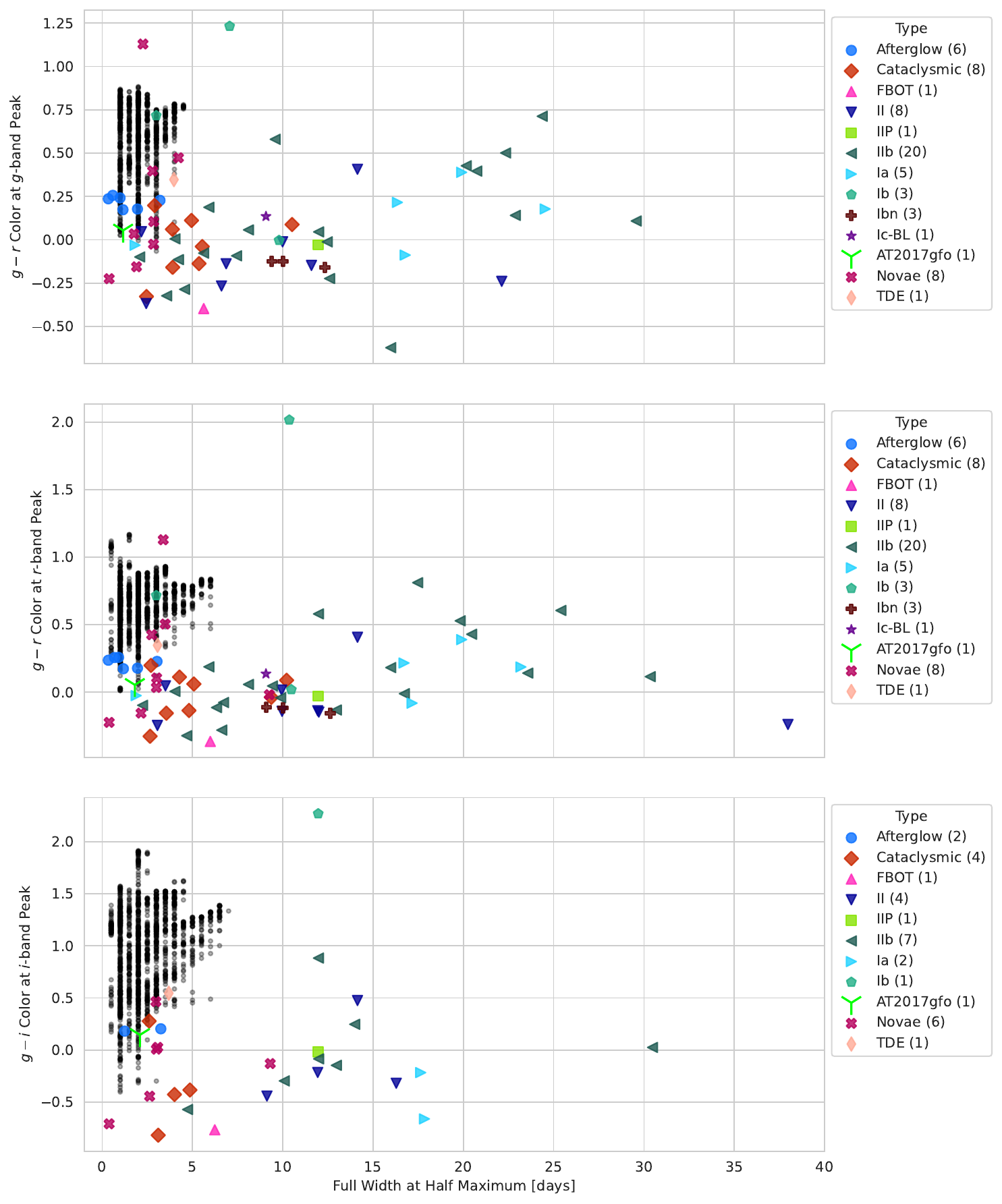}
    \caption{Full width at half maximum vs the peak color in each filter. These are both metrics that can be calculated without a distance estimate, and they demonstrate a fair degree of separation for most classes of transients with the exception of GRB afterglows.}
    \label{fig:fwhm-vs-color}
\end{figure}

Given that the color and FWHM show the most separation of the metrics tested, it is natural to compare them against one another. This is especially true given that. in the local universe, these are both metrics that are not strongly dependent on knowing the distance to an object \textit{a priori}. Figure \ref{fig:fwhm-vs-color} shows that there is a fair amount of separation between the model grid and most transient types, though the various SNe types referenced in Section \ref{sec:fwhm} still demonstrate the most separation. 

\subsection{Metric Distance}\label{sec:metric-distance}
\begin{figure}[h]
    \centering
    \includegraphics[width=0.7\linewidth]{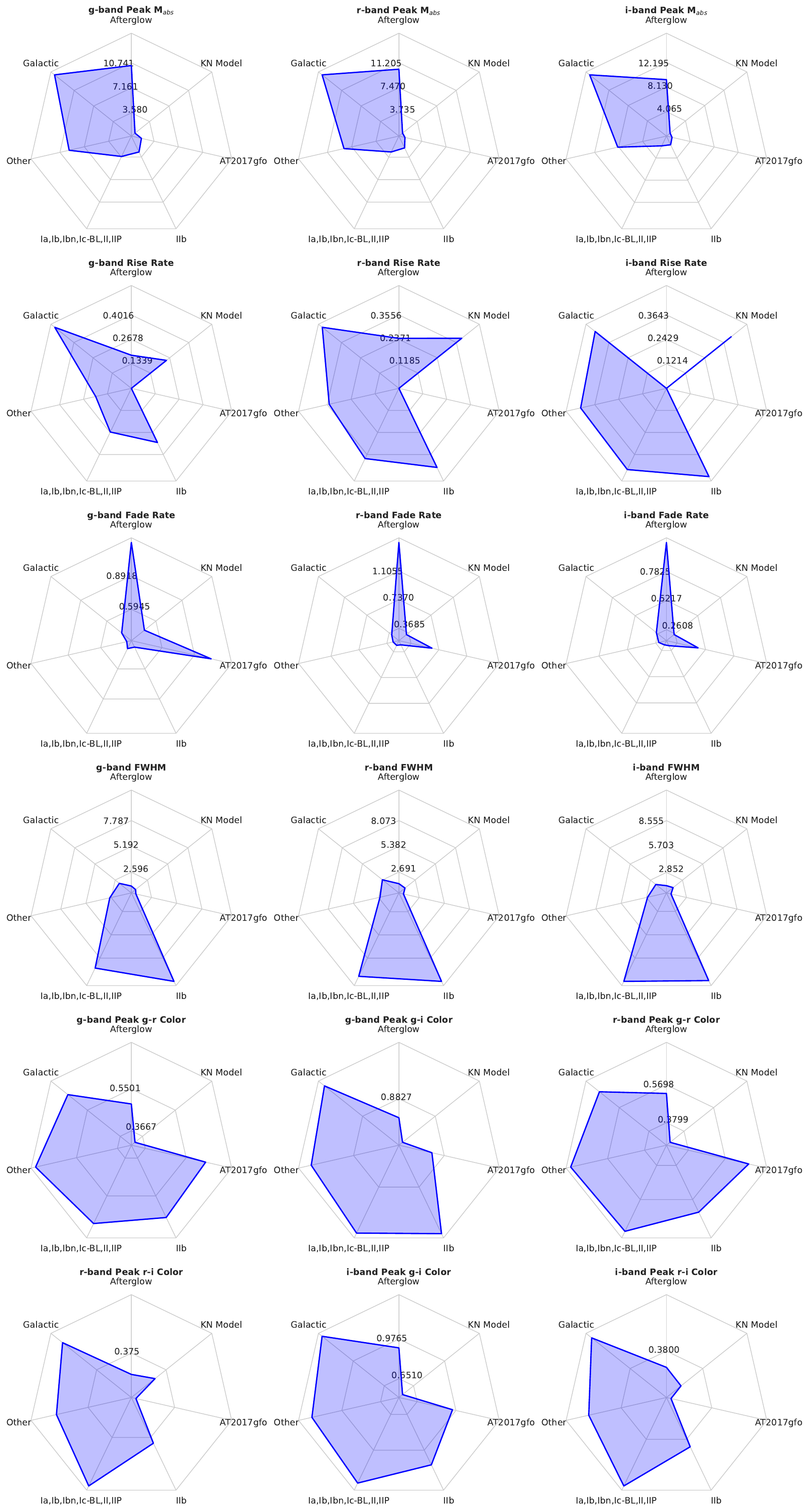}
    \caption{A collection of radar plots that correspond to the average difference between the values of each metric for a given category of transient and the values calculated for the KN model grid. If a metric cannot be computed for a specific category, the spoke for that category in that radar plot are shown as having a value of zero. For example, there is not a computed value for the rise rate of AT2017gfo, as it was only observed during the fading period.}
    \label{fig:radar}
\end{figure}

In order to compare performance between different metrics, it is necessary to identify a quantitative measure of the imposters' proximity to the model grid. One method of evaluating this is to calculate the average distance between the metric values for each class of imposter and all points in the model grid

\begin{equation}
    d = \frac{\sum_{i=0}^{n_{type}} \sum_{j=0}^{n_{grid}} | x_i - x_j |}{n_{type} \times n_{grid}}
\end{equation}

where $n_{type}$ is the number of objects contained within a given transient grouping, $n_{grid}$ is the number of simulated KNe in the model grid, $x_i$ is the metric value for a given transient, and $x_j$ is the metric value for a KNe in the model grid.

A larger value would then indicate that metric discriminates more between the KN model grid and a given class of transient.

For visual clarity, we condense the imposters into 5 groups

\begin{itemize}
    \item GRB afterglows
    \item IIb SNe
    \item Misc SNe (Ia, Ib, Ibn, II, IIp)
    \item Galactic transients (CVs, novae)
    \item Other transients (FBOTs, TDEs)
\end{itemize}

Additionally, we calculate the metric distance for AT2017gfo and a subsample of 100 lightcurves from the KN model grid. Calculating the metric distances for AT2017gfo compares the performance of the model against a well-observed KN, and the subsample from the model grid effectively measures the baseline self-similarity of the grid. This can be roughly interpreted as the baseline value of that metric distance; values below this might not be significant enough to meaningfully discriminate between an imposter and a genuine candidate KN.

Figure \ref{fig:radar} shows radar plots for all of the metrics discussed in the preceding subsections. Take the first row, which corresponds to the peak luminosity, as a test case. The KN model has a fairly narrow range in terms of peak luminosity, so the self similarity value is almost zero. Across all filters, galactic sources show the highest value, which is logical given that these transients are known to have much lower peak absolute luminosities in comparison to KNe and other extragalactic transients; however, they can be a contaminant due to their relative proximity resulting in a high apparent magnitude. One should also note that, while it is fairly small, AT2017gfo has a value slightly above that of the KN model baseline, which is most pronounced in the \textit{g}-band.

\subsubsection{Rise and Fade Rates}
The second row of Figure \ref{fig:radar} shows the rise rate across all filters. As shown in Figure \ref{fig:rise-vs-peak}, there is a fairly large range of plausible rise rates for the KN model grid, and, as a result, the baseline value is fairly high. Though there is only one GRB afterglow with an observed rise (only in the \textit{g}-band and \textit{r}-band, however), it has a metric value slightly below that of the model grid, indicating that rise rate may not be a strong filter for GRB afterglows. Again, galactic sources are the strongest outliers. The IIb SNe group and the general SNe group both exceed the model grid baseline, with the former edging the latter out slightly across all filters.  However, it should be noted that, under realistic observing conditions, it's more likely that the rise will not be able to be calculated with the same precision, so it's possible this metric may not be as performant in some cases. It's not possible to compare the model against AT2017gfo because there is not observational data prior to the peak for any filter. 

The fade rate plots on the third row show this metric is most suited to identifying GRB afterglows, which makes sense given their characteristic fast fade. In addition to this, AT2017gfo is the only other category that is notably above the baseline model grid value. Referring back to Figure \ref{fig:fade-vs-peak}, AT2017gfo is noticeably outside the bounds of the model grid in the \textit{g}-band but falls within the upper boundary of the other two filters. Despite this, the fade metric distance is still somewhat high in all filters for AT2017gfo. The \textit{g}-band being the most pronounced of the three can be explained by the aforementioned excess fade rate, but the other two filters could be explained by the gap in populations between faster and slower fading KNe in the grid, which would exaggerate the metric distance somewhat

\subsubsection{Full Width at Half Maximum}
The fourth row shows the FWHM for each filter. As shown in Figures \ref{fig:fwhm-vs-peak} and \ref{fig:fwhm-vs-color}, the range of FWHM values is fairly narrow for the model grid; correspondingly, the baseline metric distance value is relatively low. The transient classes demonstrating the greatest distance from the baseline are both categories of SNe, with IIb SNe being edged out by the other collected SNe category. SNe occur on a longer timescale in comparison to KNe, and this is reflected in Figure \ref{fig:fwhm-vs-peak}. The galactic and miscellaneous sources are also slightly above the baseline value, but much less so. 

\subsubsection{Color}
Finally, the fifth and sixth rows in Figure \ref{fig:radar} show the various color metrics. The color values of the KN model grid is fairly tight in its spread in Figure \ref{fig:color-vs-peak}, which is largely reflected in the baseline value in the radar plots being somewhat low. With the exception of the \textit{r}-band peak \textit{r}-\textit{i} color, all classes of transient exceed the baseline. Generally, the afterglow category is the closest to the baseline in each of the metrics. Interestingly, AT2017gfo exceeds the baseline in four of the six metrics, with the only exceptions being the \textit{r}-band peak \textit{r}-\textit{i} color and the \textit{i}-band peak \textit{r}-\textit{i} color. Referring back to Figure \ref{fig:color-vs-peak}, AT2017gfo is not necessarily outside the bounds of the model grid distribution, but the spread of the model grid is somewhat large. This suggests this approach may not be as effective in cases where the metric space for the baseline is fairly wide.

\section{Conclusions}\label{sec:conclusions}

There is a subset of transients, specifically novae and cataclysmic variables, that can be discriminated from the remaining population of transients based on their peak absolute luminosity in all bands, whereas most of the other objects have peak absolute luminosities within a couple of magnitudes of the model grid. This subgroup of objects, novae, classical novae, and cataclysmic variables, share the common feature of their relative proximity (z$<$0.001) compared to the rest of the sample. Additionally, across all metrics besides their absolute peak luminosity, these nearby objects have values similar to those of the KN model grid.

Holistically, it is evident that different metrics filter out certain transient types better than others - there is no one ``silver bullet" metric that definitively separates all transient types from the KN model grid. One of the qualitatively better metrics, the FWHM, has a significant trade-off; the FWHM can only be measured after an object has faded to at least half of its peak brightness. Given how quickly KNe fade beyond detectable levels for most instruments, this limits ToO follow-up to a fairly brief window of observation. Even measuring the color at the peak is a metric that limits the ToO observation window to the fade of a fast-fading transient.

While there is variance in the qualitative performance of each metric, most of the metrics discussed in this paper show a mild discriminatory utility; using these metrics in tandem with some sort of weighting could be used to calculate a second order metric, along the lines of the metric distance approach discussed in Section \ref{sec:metric-distance}, to indicate the likelihood of a transient being a KN candidate. This would return results on a timescale shorter than inference-based methods that use tools like NMMA \citep{10.1093/mnras/stae1164}.

There is evidence that this strategy has merit. The Bright Transient Survey (BTS) implemented a convolutional neural network to identify objects of interest, and they found that this automated tool triggered follow-up observations upwards of 7 hours earlier than manual inspection \cite{2023arXiv230707618R}. Given the nature of ground-based observing, 7 hours can often be the difference between a follow-up observation on the same night and having to wait until the next evening.

The authors welcome additional work in this area and in exploring additional evaluation metrics; of particular benefit would be the identification of a metric that can discriminate between imposters and candidate KNe during their rise. The data used for this analysis are publicly available on the \href{https://github.com/tylerbarna/IIb-or-not-IIIb}{IIb or not IIb GitHub repository}.

\section*{Acknowledgments}
The authors thank Theo Bonzi for his work on developing the GP model used to interpolate the object sample.

M.W.C acknowledges support from the National Science Foundation with grant numbers PHY-2117997, PHY-2308862 and PHY-2409481. M.B. acknowledges the Department of Physics and Earth Science of the University of Ferrara for the financial support through
the FIRD 2024 grant. 

Tyler Barna is supported through the University of Minnesota Data Science in Multi-Messenger Astrophysics (DSMMA) program by the National Science Foundation with grant number NRT-1922512. Any opinions, findings, and conclusions or recommendations expressed in this material are those of the author(s) and do not necessarily reflect the views of the National Science Foundation. 

Based on observations obtained with the Samuel Oschin Telescope 48-inch and the 60-inch Telescope at the Palomar Observatory as part of the Zwicky Transient Facility project. ZTF is supported by the National Science Foundation under Grants No. AST-1440341 and AST-2034437 and a collaboration including current partners Caltech, IPAC, the Weizmann Institute of Science, the Oskar Klein Center at Stockholm University, the University of Maryland, Deutsches Elektronen-Synchrotron and Humboldt University, the TANGO Consortium of Taiwan, the University of Wisconsin at Milwaukee, Trinity College Dublin, Lawrence Livermore National Laboratories, IN2P3, University of Warwick, Ruhr University Bochum, Northwestern University and former partners the University of Washington, Los Alamos National Laboratories, and Lawrence Berkeley National Laboratories. Operations are conducted by COO, IPAC, and UW.

The ZTF forced-photometry service was funded under the Heising-Simons Foundation grant \#12540303 (PI: Graham).

\newpage
\appendix

\section{A Glossary of Transients}\label{sec:classes}

There are a number of classes of transient present in this sample of KN imposters. Each of these classes have unique mechanisms and properties with rich fields of study. Here, we provide brief summaries of each of the classes included in our refined sample.

\subsection{GRB Afterglows}
GRB afterglows (referred to as afterglows for much of this paper) occur as a result of the shock originating from a GRB interacting with the interstellar medium in the form of synchrotron emission \citep{1998ApJ...497L..17S}. Like KNe, afterglows decline rapidly \citep{1999ApJ...525..737R}. 

CBCs give rise to both a KN and a short GRB (sGRB) afterglow, as was observed with GW170817 and its counterparts \citep{2017ApJ...848L..12A}. In these cases, it can be difficult to disentangle the two signals from each other given the greater brightness of afterglows \citep{2022ApJ...938..147Z}, though multi-wavelength observations can mitigate this challenge \citep{2024arXiv240907539W}. 

\subsection{Cataclysmic Variables}
Cataclysmic variables (CVs), labeled as ``Cataclysmic" in most of our figures, are binary star systems consisting of a white dwarf (WD) that accretes matter from its companion. This accretion can lead to outbursts in the form of classical novae, recurrent novae, dwarf novae, and nova-like objects, among other more exotic variants \citep{1976ARA&A..14..119R, 1999PASP..111.1281S}. Despite the``recurrent" subclass, all types of novae are technically periodic in nature on varying timescales; even classical novae are thought to be periodic, albeit on the time scale of thousands of years \citep{1978MNRAS.183..515B}. Some of the CVs in this sample could very well be novae. For whatever reason, they were classified more generically; additional photometric and spectroscopic observation could clarify their subclass \citep{2008ApJ...683.1006K}. 

\subsection{Novae}
As a subclass of CV, novae describe instances wherein the white dwarf (WD) accreting mass leads to a outburst in the form of thermonuclear runaway  on the surface of the WD \citep{1972ApJ...176..169S}. Depending on emission lines observed in their spectroscopy, they can be divided into ``He/N" and ``Fe II" novae \citep{2012AJ....144...98W}.

In addition to classical novae, there are a number of subclasses, such as dwarf novae, which are driven by an increase in luminosity in the accretion disk \citep{1974MNRAS.168..235W, 1996PASP..108...39O} rather than on the surface of the WD. Dwarf novae tend to be dimmer than classical novae. 

In general, novae are less luminous that many other classes of transient (hence the distinction between novae and ``super"-novae \citep{1934CoMtW...3...79B, 2001AAS...199.1501O}, but they occur at a relatively higher rate within the galaxy \citep{2017ApJ...834..196S}. With limited photometric observation and no spectroscopy, it is possible for a faint galactic nova to initially be mistaken for a brighter event occurring at a greater distance.

\subsection{Fast Blue Optical Transients}
Fast blue optical transients (FBOTs) are a somewhat new class of transient characterized by a blue color and a rapid rise and decay; there are a number of proposed models for their progenitors \citep{2020ApJ...903...66L, 2020ApJ...904..155A, 10.1093/mnrasl/slad145, 2020ApJ...903...66L, 2022MNRAS.513.3810G, 2022RAA....22e5010S}.

The most famous example of an FBOT is AT2018cow, also referred to as ``The Cow" \citep{2020ApJ...903...66L}. The Cow was an extremely rapid transient that exceeded the peak luminosity of a super-luminous supernovae (SLSN). The FBOT in our sample, ZTF23aaeozpp/AT2023fhn, also known as ``The Finch," was similar to AT2018cow, albeit with a significant separation from its host galaxy \citep{10.1093/mnrasl/slad145}.

FBOTs are rare events, with some estimates placing them at around $0.1\%$ of the local core-collapse supernova (CCSN) rate \citep{2023ApJ...949..120H}. While there have been upper limit estimates of the rate of KNe \citep{2020ApJ...904..155A}, the actual number of FBOTs observed exceeds that of KNe; this, in tandem with their rapid evolution, make them potential KNe imposters.

\subsection{Tidal Disruption Events}
Tidal disruption events (TDEs) occur when a star orbiting a supermassive black hole (SMBH) is drawn so close to the SMBH that it is pulled apart by tidal forces. This phenomenon was first theorized in relation to quasi-stellar sources (QSOs) \citep{1975Natur.254..295H}. While the first observations of TDEs occurred in the x-ray regime \citep{1988ApJ...325L..25P, 2002AJ....124.1308D}, TDEs are observable across the EM spectrum depending on the viewing angle \citep{2018ApJ...859L..20D}.   A number of TDEs have been detected by ZTF \citep{2021ApJ...908....4V}, making them a potential contaminant, but their presence could be mitigated with galaxy correlation given their presence at the center of galactic nuclei and their higher peak luminosity in comparison to KNe \citep{2011MNRAS.410..359L}.

Some TDEs may however be found to be offset from their host galaxy nucleus. Three such cases have been reported in the literature \citep{2018NatAs...2..656L, 2025arXiv250109580J, 2025ApJ...985L..48Y}, two of which were observed primarily in the X-rays \citep{2018NatAs...2..656L, 2025arXiv250109580J}. In the case of the optically-discovered off-nuclear event AT2024tvd \citep{2025ApJ...985L..48Y}, the transient reached maximum light ($M_r ~ -19$ mag) in $\sim20$ days in the rest frame, then faded slowly. Its peak brightness and light curve are therefore distinct from any plausible kilonova models. Because of their rarity and light curve behavior, we expect contamination from off-nuclear TDEs to be negligible. 

\subsection{Type Ia Supernovae}
There are two primary methods of delineating the various classes of SN, the first being by the presence of absorption lines of specific elements in their spectra and the second being their explosion mechanism. The former was initially described in \cite{1941PASP...53..224M} and has been adopted as the primary classification method and has been expanded in greater detail \citep{1993Ap&SS.202..215D}. Broadly, Type I SNe are defined by the presence of a hydrogen absorption line, whereas Type II SNe do demonstrate a hydrogen absorption line.

Type Ia supernovae (SNe Ia) are differentiated by the presence of a silicon line in their spectra \citep{1997ARA&A..35..309F}. They are also the only class of SN that undergoes thermonuclear runaway for its explosion mechanism. This occurs when a WD in a binary system accretes enough mass to hit the Chandrasekhar limit, which leads to the thermonuclear runaway \citep{1931ApJ....74...81C, 1971ApJ...170..299B}.
Due to this roughly similar explosion mass, SNe Ia are understood to have standardizable peak luminosities \citep{1993ApJ...413L.105P}, though there are a subset of SNe Ia that are under-luminous, which are referred to as SNe Iax \citep{2013ApJ...767...57F}.

The exact nature of the progenitor systems of SNe Ia is not fully understood, with a number of different models in contention \citep{ 2014ARA&A..52..107M, 2025A&ARv..33....1R}. 

\subsection{Type Ib Supernovae}
Like SNe Ia, SNe Ib are categorized as Type I SNe due to the paucity of hydrogen in their spectra, but they demonstrate a helium line that is not present in SNe Ia and generally lack a strong silicon absorption line \citep{1997ARA&A..35..309F}. SNe Ib are driven by core collapse (CC), which occurs when the core of the star is no longer able to counter the gravitational forces of its own mass \citep{clocchiatti1997light}. CC SNe progenitors are massive stars, generally of the order 8 $M_\odot$ at minimum \citep{Heger_2003, 2021MNRAS.506..781D}, with some debate as to the upper mass limit \citep{Smartt_2009, Smartt_2015, 2016ApJ...821...38S, 2018A&A...613A..35K, 2021MNRAS.506..781D}.

Unlike SNe Ic, SNe Ib possess a helium envelope \citep{modjaz2009shock}. The early lightcurves of SNe Ib are driven by the expansion of this stripped envelope (SE), later progressing into the decay of $^{56}$Ni. While the lightcurves of SNe Ib may initially evolve in a manner similar to SNe Ia, SNe Ib demonstrate a longer fade time after their peak as well as a redder color \citep{2012MNRAS.424.2139D}.

\subsection{Type Ibn Supernovae}
As a class of SN, SNe Ibn are distinguished not by their explosion mechanism but by the environment in which they occur. SNe Ibn are SNe Ib that occur in the presence of a circumstellar medium (CSM) \citep{2017hsn..book..403S}. This CSM has a shock interaction with the SN ejecta, which can result in unexpected behavior such as reddening \citep{2015MNRAS.454.4293P} and demonstrating overlap with SNe IIb in some instances \citep{2020MNRAS.499.1450P}.

\subsection{Type II Supernovae}

Like SNe Ib, SNe II occur as a result of the CC of massive stars, but they differ in that their red supergiant progenitors retain their hydrogen envelope and demonstrate a hydrogen absorption line in their spectra \citep{chevalier1994emission, filippenko1997optical}.
There are a number of subclasses of SNe II, of which two are described below. 

The two primary subclasses, IIP and IIL, are categorized based on the photometric properties of their fade following their peak, though recent studies have investigated the prospects of distinguishing between SNe II subclasses based on spectropolarimetry \citep{Nagao_2024}. Objects may be classified as a non-specific SN II depending on the extent of their observation. In our sample, we have relied upon existing formal classifications documented in the Transient Name Server (TNS) \citep{2021AAS...23742305G}, General Coordinates Network (GCN) \citep{2000AIPC..526..731B, 2023AAS...24110802S}, Lasair \citep{2019RNAAS...3...26S}, and other sources rather than evaluate the nature of included objects as SNe IIP or SNe IIL.

\subsection{Type IIP Supernovae}
SNe IIP are a subclass of SNe II defined by the presence of an extended plateau\footnote{In contrast, SNe IIL demonstrate a linear decay in luminosity following their peak \citep{2014MNRAS.445..554F}, though this subclass does not appear in our sample, potentially owing to SNe IIL occurring at a lower rate than SNe IIP \citep{2011MNRAS.412.1441L}.} 
in their photometric lightcurve following a brief fade from the peak, hence the ``P." This plateau phase lasts on the order of roughly 100 days \citep{poznanski2009improved, arcavi2012caltech}. This occurs because the outer hydrogen envelope is ionized, creating an optically thick layer that obscures radiation inside of this layer. As the ejecta expands, the ionized hydrogen recombines and becomes optically thin, and this process repeats throughout the expanding hydrogen envelope \citep{1971Ap&SS..10...28G}. This phenomenon results in a consistent release of energy until the entirety of the hydrogen envelope is recombined, which is observable as a characteristic plateau \citep{1994ApJ...430..300E}.
SNe IIP can demonstrate a range in magnitudes at which they plateau \citep{hamuy2003observed, 2006A&A...450..345K}.

\subsection{Type IIb Supernovae}

SNe IIb are a unique class of CC SNe that initially demonstrate spectra that appear similar to SNe II but later evolve to more closely resemble SNe Ib \citep{filippenko1997optical}. This occurs due to a milder stripping of the outer hydrogen layers at early times in comparison to other SE SNe, with the associated hydrogen spectral features disappearing at later times, coming to resemble the spectra of SNe Ib \citep{2019ApJ...885..130S}. A unique observational trait of SNe IIb is their double-peak feature \citep{2025arXiv250303735C}. The initial peak is powered by shock-cooling emission \citep{2016ApJ...818..111K, 2021MNRAS.507.3125A}, which later gives way to the secondary peak that is driven by the decay of $^{56}$Ni produced during the initial explosion \citep{1980ApJ...237..541A, Pellegrino_2023}. The first peak demonstrates a more rapid rise and fade than the second, but both peaks fade faster than something like a SN IIP.
SNe IIb represent the largest subclass of observed SE SNe \citep{2011MNRAS.412.1473L, 2017PASP..129e4201S}, though they are generally overrepresented in our sample relative to SNe IIP, which are the most observed subclass of SNe generally \citep{2017PASP..129e4201S}, suggesting they are a major contaminant in KNe searches.

\section{Transient List}
\begin{longtable}[h]{|l|l|l|l|l|}
\caption{Imposter Transients} \label{tab:objects} \\ \hline
\textbf{Object} & \textbf{First Observation} & \textbf{Type} & \textbf{Redshift} & \textbf{Citation} \\ \hline
\endfirsthead
\multicolumn{4}{c}
{\tablename\ \thetable\ -- \textit{Continued from previous page}} \\
\hline
\textbf{Object} & \textbf{First Observation} & \textbf{Type} & \textbf{Redshift} & \textbf{Citation} \\ \hline
\endhead
\hline \multicolumn{4}{r}{\textit{Continued on next page}} \\
\endfoot
\hline
\endlastfoot
ZTF18aakuewf & 2018-04-18 & Ibn & 0.0636 & \cite{ZTF18aakuewf} \\ \hline
ZTF18aalrxas & 2018-04-19 & IIb & 0.0582 & \cite{ZTF18aalrxas} \\ \hline
ZTF18abffyqp & 2018-07-08 & II & 0.031 & \cite{ZTF18abffyqp} \\ \hline
ZTF18abvkmgw & 2018-09-12 & Ib & 0.03847 & \cite{ZTF18abvkmgw} \\ \hline
ZTF18abwkrbl & 2018-09-15 & IIb & 0.00999 & \cite{ZTF18abwkrbl} \\ \hline
ZTF19aanbpus & 2020-10-09 & IIb & 0.009969 & \cite{ZTF19aanbpus} \\ \hline
ZTF19aapfmki & 2019-04-10 & Ibn & 0.05469 & \cite{ZTF19aapfmki} \\ \hline
ZTF19aatesgp & 2019-04-29 & Ib & 0.0043 & \cite{ZTF19aatesgp} \\ \hline
ZTF19abacxod & 2019-06-19 & II & 0.01765 & \cite{ZTF19abacxod} \\ \hline
ZTF19abxjrge & 2019-09-06 & IIb & 0.021705 & \cite{ZTF19abxjrge} \\ \hline
ZTF19abxtcio & 2019-09-07 & IIb & 0.0155 & \cite{ZTF19abxtcio} \\ \hline
ZTF19abyjzvd & 2019-09-11 & Ibn & 0.1353 & \cite{ZTF19abyjzvd} \\ \hline
ZTF19acbumks & 2019-10-01 & IIb & 0.031 & \cite{ZTF19acbumks} \\ \hline
ZTF19accjfgv & 2019-10-03 & IIb & 0.027 & \cite{ZTF19accjfgv} \\ \hline
ZTF20aahfqpm & 2020-01-23 & IIb & 0.03113 & \cite{ZTF20aahfqpm} \\ \hline
ZTF20aajnksq & 2020-01-28 & Afterglow & 2.9 & \cite{ZTF20aajnksq} \\ \hline
ZTF20aaxhzhc & 2020-04-29 & IIb & 0.037 & \cite{ZTF20aaxhzhc} \\ \hline
ZTF20aayrobw & 2020-05-11 & II & 0.061 & \cite{ZTF20aayrobw} \\ \hline
ZTF20aazchcq & 2020-05-10 & II & 0.03788 & \cite{ZTF20aazchcq} \\ \hline
ZTF20abstsxb & 2020-08-15 & Ia & 0.094 & \cite{ZTF20abstsxb} \\ \hline
ZTF20aburywx & 2020-08-19 & IIb & 0.0313 & \cite{ZTF20aburywx} \\ \hline
ZTF20abwzqzo & 2020-08-26 & IIb & 0.023033 & \cite{ZTF20abwzqzo} \\ \hline
ZTF20acgigfo & 2020-10-02 & Novae & 0.00017 & \cite{ZTF20acgigfo} \\ \hline
ZTF20acgiglu & 2020-10-02 & IIb & 0.026728 & \cite{ZTF20acgiglu} \\ \hline
ZTF20acigusw & 2020-10-12 & II & 0.062 & \cite{ZTF20acigusw} \\ \hline
ZTF20aclfmwn & 2020-10-19 & IIb & 0.038 & \cite{ZTF20aclfmwn} \\ \hline
ZTF20acozryr & 2020-11-04 & Afterglow & 1.105 & \cite{ZTF20acozryr} \\ \hline
ZTF20acqntkr & 2020-11-14 & Ia & 0.015661 & \cite{ZTF20acqntkr} \\ \hline
ZTF20actqnhg & 2020-11-25 & IIb & 0.031 & \cite{ZTF20actqnhg} \\ \hline
ZTF21aaabrpu & 2021-01-01 & IIb & 0.027172 & \cite{ZTF21aaabrpu} \\ \hline
ZTF21aaabwfu & 2021-01-01 & IIb & 0.010929 & \cite{ZTF21aaabwfu} \\ \hline
ZTF21aabxjqr & 2021-01-07 & IIb & 0.03317803 & \cite{ZTF21aabxjqr} \\ \hline
ZTF21aagwbjr & 2021-02-05 & Afterglow & 0.876 & \cite{ZTF21aagwbjr} \\ \hline
ZTF21aapkbav & 2021-03-17 & IIb & 0.036 & \cite{ZTF21aapkbav} \\ \hline
ZTF21aaqzrat & 2021-03-30 & Cataclysmic & 0.0 & \cite{ZTF21aaqzrat} \\ \hline
ZTF21abbvvmf & 2021-05-18 & IIb & 0.010945 & \cite{ZTF21abbvvmf} \\ \hline
ZTF21abexegc & 2021-06-08 & Cataclysmic & 0.0 & \cite{ZTF21abexegc} \\ \hline
ZTF21abfmpwn & 2021-06-11 & Afterglow & 1.1345 & \cite{ZTF21abfmpwn} \\ \hline
ZTF21aciosfu & 2021-10-22 & Cataclysmic & 0.0 & \cite{ZTF21aciosfu} \\ \hline
ZTF22aaajecp & 2022-02-12 & TDE & 1.1933 & \cite{ZTF22aaajecp} \\ \hline
ZTF22aabjpxh & 2022-02-19 & Afterglow & 0.293 & \cite{ZTF22aabjpxh} \\ \hline
ZTF22aahgvlx & 2022-04-24 & IIb & 0.056 & \cite{ZTF22aahgvlx} \\ \hline
ZTF22aaibbvy & 2022-04-28 & Cataclysmic & 0.0 & \cite{ZTF22aaibbvy} \\ \hline
ZTF22aajrrzz & 2022-05-11 & Ia & 0.034129 & \cite{ZTF22aajrrzz} \\ \hline
ZTF22aayluxo & 2022-08-03 & Cataclysmic & 0.0 & \cite{ZTF22aayluxo} \\ \hline
ZTF22aazmooy & 2022-08-09 & Novae & 0.00017 & \cite{ZTF22aazmooy} \\ \hline
ZTF22abfjnpj & 2022-09-05 & Cataclysmic & 0.0 & \cite{ZTF22abfjnpj} \\ \hline
ZTF22abfxmpc & 2022-09-17 & Novae & 0.0 & \cite{ZTF22abfxmpc} \\ \hline
ZTF22abijszk & 2022-09-26 & Novae & 0.001 & \cite{ZTF22abijszk} \\ \hline
ZTF22abmsaxp & 2022-10-12 & II & 0.013403 & \cite{ZTF22abmsaxp} \\ \hline
ZTF22abuzpzz & 2022-11-18 & Cataclysmic & 0.0 & \cite{ZTF22abuzpzz} \\ \hline
ZTF23aadhssd & 2023-03-17 & Cataclysmic & 0.0 & \cite{ZTF23aadhssd} \\ \hline
ZTF23aaemgsd & 2023-04-09 & Ia & 0.023 & \cite{ZTF23aaemgsd} \\ \hline
ZTF23aaeozpp & 2023-04-10 & FBOT & 0.24 & \cite{ZTF23aaeozpp} \\ \hline
ZTF23aaikakr & 2023-04-29 & Cataclysmic & 0.0 & \cite{ZTF23aaikakr} \\ \hline
ZTF23aajadma & 2023-05-09 & Cataclysmic & 0.0 & \cite{ZTF23aajadma} \\ \hline
ZTF23aaoohpy & 2023-06-18 & Afterglow & 1.027 & \cite{ZTF23aaoohpy} \\ \hline
ZTF23aarlxdf & 2023-07-12 & Novae & 0.0 & \cite{ZTF23aarlxdf} \\ \hline
ZTF23aaxeacr & 2023-08-13 & Afterglow & 0.36 & \cite{ZTF23aaxeacr} \\ \hline
ZTF23aaxzvrr & 2023-08-15 & Novae & 0.0 & \cite{ZTF23aaxzvrr} \\ \hline
ZTF23abgsmsg & 2023-09-25 & Novae & 0.0 & \cite{ZTF23abgsmsg} \\ \hline
ZTF23abjwgre & 2023-10-14 & IIP & 0.0231 & \cite{ZTF23abjwgre} \\ \hline
ZTF23abnpdod & 2023-10-28 & Cataclysmic & 0.0 & \cite{ZTF23abnpdod} \\ \hline
ZTF23abobwsd & 2023-11-01 & IIb & 0.024 & \cite{ZTF23abobwsd} \\ \hline
ZTF23absbqun & 2023-12-07 & Ib & 0.0101 & \cite{ZTF23absbqun} \\ \hline
ZTF23abtycgb & 2023-12-14 & II & 0.020127 & \cite{ZTF23abtycgb} \\ \hline
ZTF24aascytf & 2024-06-15 & II & 0.038 & \cite{ZTF24aascytf} \\ \hline
ZTF24abvevzs & 2024-12-02 & IIb & 0.035 & \cite{ZTF24abvevzs} \\ \hline

\end{longtable}

\bibliography{bibliography}{}
\bibliographystyle{aasjournal}

\end{document}